\documentclass{article}

\usepackage{arxiv}
\usepackage[utf8]{inputenc}
\usepackage[T1]{fontenc}
\usepackage{hyperref}
\usepackage{url}
\usepackage{booktabs}
\usepackage[backend=biber,style=numeric-comp,sorting=none]{biblatex}
\usepackage{amsfonts}
\usepackage{nicefrac}
\usepackage{microtype}
\usepackage{graphicx}
\usepackage{subfig}
\usepackage{float}
\usepackage{geometry}
\usepackage{color, colortbl}
\usepackage{rotating}
\usepackage{subfig}
\usepackage{tikz}
\usepackage[all]{xy}
\usetikzlibrary{positioning} 
\usetikzlibrary{arrows,decorations.pathmorphing,shapes} 
\usepackage{tikz-cd}
\usepackage{xcolor}
\usepackage{animate}
\usetikzlibrary{shapes.geometric, arrows}
\usepackage{amsmath}
\usepackage[export]{adjustbox}
\usepackage[ruled,vlined]{algorithm2e}

\usepackage{mathtools}

\DeclareMathOperator*{\argminA}{arg\,min}

\title{Psychophysics of Artificial Neural Networks \\ Questions Classical Hue Cancellation Experiments}

\author{
    Jorge Vila-Tomás\\
    Image Processing Laboratory\\
    Universitat de València\\
    Valencia, 46980, Spain  \\
  \texttt{jorge.vila-tomas@uv.es} \\
\And
    Pablo Hernández-Cámara\\
    Image Processing Laboratory\\
    Universitat de València\\
    Valencia, 46980, Spain  \\
  \texttt{pablo.hernandez-camara@uv.es} \\
\And
    Jesús Malo\\
    Image Processing Laboratory\\
    Universitat de València\\
    Valencia, 46980, Spain  \\
  \texttt{jesus.malo@uv.es} \\
}

\begin{document}

\definecolor {processblue}{cmyk}{0.96,0,0,0}
\maketitle

\begin{abstract}

We show that classical hue cancellation experiments lead to human-like opponent curves even if the task is done by trivial (\emph{identity}) artificial networks.
Specifically, human-like opponent spectral sensitivities always emerge in artificial networks as long as (i)~the \emph{retina} converts the input radiation into any tristimulus-like representation, 
and (ii)~the post-retinal \emph{network} solves the standard hue cancellation task, e.g. the network looks for the weights of the cancelling lights so that every monochromatic stimulus plus the weighted cancelling lights match a grey reference in the (arbitrary) color representation used by the network.
In fact, the specific cancellation lights (and not the network architecture) are key to obtain human-like curves: results show that the classical choice of the lights is the one that leads to the best (more human-like) result, and any other choices lead to progressively different spectral sensitivities.
We show this in two ways:
through \emph{artificial psychophysics} using a range of networks with different architectures and a range of cancellation lights, and through a \emph{change-of-basis theoretical analogy} of the experiments. This suggests that the opponent curves of the classical experiment are just a by-product of the front-end photoreceptors and of a very specific experimental choice but
they do not inform about the downstream color representation.
In fact, the architecture of the post-retinal network (signal recombination or internal color space) seems irrelevant for the emergence of the curves in the classical experiment.
This result in artificial networks questions the conventional interpretation of the classical result in humans by Jameson and Hurvich.

\end{abstract}


\section{Introduction}

The classical hue cancellation experiments~\cite{Jameson55_1,Jameson57} 
are usually considered as the first psychophysical quantification of Hering's intuition on opponent color coding in the human brain~\cite{Fairchild13,Stockman11,Shevell04}. As an example, an influential textbook on visual neuroscience~\cite{Wandell95} introduces hue cancellation as follows: \emph{"Several experimental observations, beginning in the mid-1950s, catapulted opponent-colors theory from a special-purpose model, known only to color specialists, to a central idea in Vision Science. The first was a behavioral experiment that defined a procedure for measuring opponent-colors, the hue cancellation experiment. By providing a method of quantifying the opponent-colors insight, Hurvich and Jameson made the idea accessible to other scientists, opening a major line of inquiry."}

\begin{figure}[t!]
\begin{centering}
\hspace{-1cm}
\includegraphics[width=17cm]{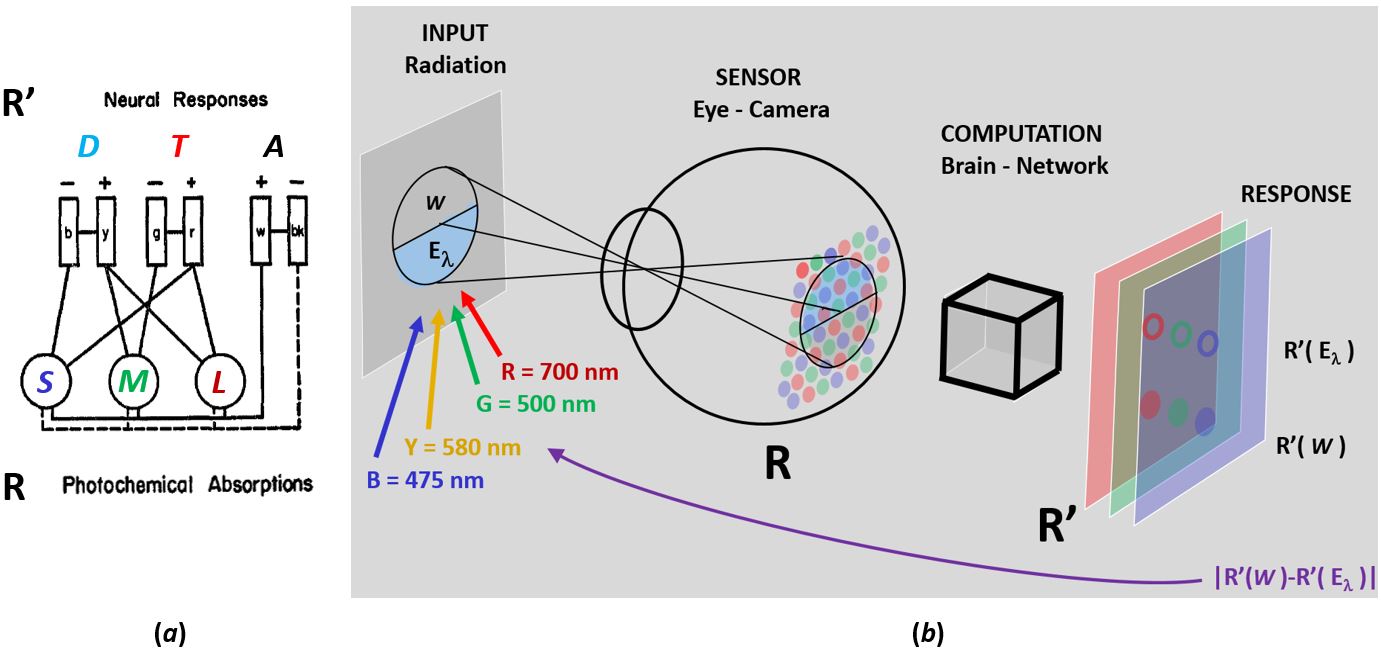}
\end{centering}
\caption{\textbf{(a)} Elements of the competing theories of Young-Helmholtz vs Hering, and \textbf{(b)} Learning process to get the weights that cancel the hue of certain monochromatic stimulus of wavelength $\lambda$. Following the original diagram in \cite{Jameson57}, \textbf{Figure 1.a} displays the sensors of the Young-Helmholtz theory, with all-positive sensitivities tuned to \emph{Long}, \emph{Medium}, and \emph{Short} (LMS) wavelengths, and a possible architecture of a network that would lead to the sensors of the Hering theory: two chromatic sensors with opponent sensitivities, the \emph{Tritanopic} sensor (T) tuned to red-green and the \emph{Deuteranopic} sensor (D) tuned to yellow-blue, together with an Achromatic sensor (A) with a wide all-positive sensitivity. 
\textbf{Figure 1.b} illustrates the hue cancellation experiment: the (natural or artificial) observer \emph{looks for} the weights of the spectral cancelling lights so that a mixture of these cancellation stimuli with the original monochromatic input matches a grey reference (a stimulus with no hue). In this setting, hue cancellation reduces to distance minimization between the responses $R'$ to the white and to the considered $\lambda$ plus the weighted cancelling lights.\\
\textbf{The question} is whether this search of the weights reveals something about the computation or architecture of the \emph{brain-network} module in Fig. 1.b that transforms $R$ into $R'$, or about the nature of the inner color representation $R'$.
\label{scheme}}
\end{figure}

The scientific question to be solved by the \emph{hue cancellation experiment} is about the post-retinal neural architecture, or  recombination of color signals after photodetection. This is illustrated by Fig.~\ref{scheme}.a, based on the original diagram in~\cite{Jameson57}. The authors confront the Young-Helmholtz trichromatic theories of color vision with the qualitative opponent theory of Hering. They propose an architecture to get the Achromatic, Tritanopic (red-green) and Deuteranopic (yellow-blue) sensors (ATD) from the front-end photoreceptors tuned to Long, Medium, and Short (LMS) wavelengths, and hue cancellation would be the tool to quantify the spectral sensitivity of the ATD mechanisms in the proposed architecture.  

In this work we present a counter-example based on artificial networks (on automatic differentiation) 
that suggests that the results of conventional hue cancellation experiments do not provide conclusive information on the inner color representation of the system that mediates the task (the post-retinal network, black box in Fig.\ref{scheme}.b). Therefore, strictly speaking, the curves from the classical hue cancellation experiments would not be measuring the sensitivity of those ATD mechanisms. 

In particular, we show that \emph{identity networks} develop opponent red-green and yellow-blue color valence functions which are quite similar to the human curves independently of the color representation (LMS, RGB or ATD).
What we refer to as \emph{identity network} is a trivial architecture whose (3-dimensional) output is exactly the same as its (3-dimensional) input in each spatial location. 
This trivial network, which already operates in a tristimulus-related representation, (say certain standard LMS cone space~\cite{Stockman00}, or even in an arbitrary, device dependent, digital count RGB space~\cite{Psytoolbox97,Colorlab02}) may apply no opponent color coding whatsoever and still gets the human-like curves (in contrast to the specific architecture assumed in Fig.~\ref{scheme}.a). Therefore, the opponent curves that emerge do not strictly inform of the inner (eventually opponent) color representation of the post-retinal neural network. Instead, they are a by-product of the (retinal) tristimulus representation of the input radiation and of the choices in the conventional experimental setting (e.g. the wavelengths of the spectral cancellation lights). To explore in more detail this result, we perform multiple hue cancellation experiments with cancellation lights different to the classical ones and we obtain a clear dependence with the choice of the spectral cancellation lights, achieving the best human-like behaviour only in the case of the classical cancellation lights. This result is confirmed by an analysis of the hue cancellation experiment using a change-of-basis analogy.

\section{Methods: hue cancellation experiments in artificial networks}
\label{theory}

\subsection{General setting}

In this work the artificial hue cancellation experiment is a matching problem in the color representation used by the artificial network.
Take the setting represented in Fig~\ref{scheme}.b: for any arbitrary spectral input of wavelength $\lambda$, $E_\lambda$, and a grey reference, $W$, the network takes the input retinal representation of stimulus and reference, $R(E_\lambda)$ and $R(W)$, and transforms them into the inner representation $R'(E_\lambda)$ and $R'(W)$. We make no assumption of the nature of this representation $R'$. In Fig~\ref{scheme}.b $R'$ is represented by red, green and blue layers just for visualization, this does not mean we assume them to be LMS-like.
In the initial situation, when no cancelling lights are added, the distance $|R'(W) - R'(E_\lambda)|$ will have a large value. The goal in this matching problem is looking for the optimal weights $w^\star_{\lambda_c}\!(\lambda)$ of the cancelling lights that minimize the distance between the reference and the monochromatic stimulus plus the weighted cancelling lights:
\begin{equation}
    w_{\lambda_c}^\star\!(\lambda) = \argminA_{w_{\lambda_c}\!(\lambda)} \left| R'(W) - R'\left(E_\lambda \oplus \sum_{\lambda_c} w_{\lambda_c}\!(\lambda) \, E_{\lambda_c} \right) \right|
    \label{goal}
\end{equation}
where the subtraction in the distance is regular subtraction between vectors, but $\oplus$
stands for additive superposition of radiations. 
Physical superposition is always positive so, in this case, as conventionally done in color matching experiments~\cite{Stiles00}, we assume that \emph{negative} weights in the superposition to $E_\lambda$ physically mean the corresponding amount of \emph{positive} superposition to $W$.
In short, the cancellation experiment should tell us about the change of color representations, from the input space $R$ to the output $R'$. In principle, the goal function in Eq.~\ref{goal} can be applied to regular tristimulus vectors (where vector summation has perceptual meaning) but also to arbitrary, engineering-oriented device-dependent color representations such as digital counts in RGB.

The matching problem described above is just a difference minimization problem which is well suited for learning based on automatic differentiation. 
In this \emph{artificial psychophysics} setting, the network architecture of the black-box in Fig.~\ref{scheme}.b is fixed but the energy of the cancelling lights (the weights $w_{\lambda_c}$) is modified in each iteration to minimize the distance in Eq.~\ref{goal}. 

Appendix A elaborates on how to approximate monochromatic stimuli for artificial networks intended to work with restricted stimuli such as regular digital images. Appendix B elaborates on how the four individual weighting functions we get from the artificial nets, $w^\star_{\lambda_c}\!(\lambda)$, are combined into the final valence functions (that happen to be red-green and yellow-blue in the case of the conventional $\lambda_c$'s). 

\subsection{Hue cancellation with artificial networks beyond the classical setting}

This artificial simulation of the hue cancellation experiment can be applied with any architecture in the fixed network (black box in Fig.~\ref{scheme}.b) and with any choice of $\lambda_c$'s for the cancelling lights.

If human-like opponent channels emerge from the simulations even if the network does not have a biologically plausible architecture and independently of the post retinal space, this means that the result of the classical experiment cannot be interpreted as an indication of the existence of post-retinal mechanisms performing the computation suggested in Fig.~\ref{scheme}.a. 

Refutation of the conventional interpretation of the classical experiment is stronger if the emergence of opponent curves mainly happens with a particular choice of $\lambda_c$'s.
This would mean that instead of having the result because of interesting properties of the post-retinal mechanisms, it comes from a fortunate selection of the experimental setting.
For this reason it is interesting to simulate hue cancellation for a range of alternative $\lambda_c$'s different from the classical experiment.

\subsection{Differences with the experimental setting for humans}

In the original experiments with humans, the cancelling lights had the same energy and their wavelengths were slightly different for the two observers J/H: $467/475$ nm (blue), $490/500$ nm (green), $588/580$ (yellow), and $700/700$ nm (red). In all our simulations the cancelling lights always had the same initial energy and we used an equienergetic stimulus as grey reference. In simulating the classical setting, our wavelengths were the ones for observer H ($475$, $500$, $580$ and $700$ nm).
In our experiments we use (without loss of generality) quasi-monochromatic lights so that they can be properly represented in digital values to be processed by conventional artificial networks. These stimuli are defined by a narrow Gaussian spectral radiance added on top of a low-radiance equienergetic background. Appendix A shows examples of these stimuli.

In solving the distance minimization problem, the iterative variation of the weights was applied to the height of the narrow Gaussian of the quasi-monochromatic cancelling lights. These differences (cancelling wavelengths similar to the ones in the classical experiment and narrow-spectrum quasi-monochromatic stimuli) do not imply fundamental differences with the classical setting.  

Human observers in the classical experiment do not change all 4 weights at the same time, but (just for the observers convenience) they just move one at a time (judging how the complementary hue disappears) and repeat the experiment 4 times. 
This is not a fundamental difference because (at the expense of longer time per wavelength) after the "first cancellation" the observer could also cancel the remaining hue and then match the response to a grey.
Additionally, in any part of the spectrum, is the experimenter in the classical experiment who lets the observer to use "the appropriate" cancellation light. This is not a fundamental difference either because if they could look for the cancellation lights in pairs, simultaneous modification of the opponent cancellation lights would null each other and the effect would be as using a single one.

In the setting that we propose to simulate hue cancellation in artificial systems, the only difference with regard to the experiments in humans is that humans may not need an achromatic reference since they already have the concept of what an achromatic stimulus is, and hence they modify the weights of the cancellation lights to match this mental concept.
In the case of artificial systems, obtaining the concept of achromatic reference for hue cancellation is not a problem either. It could be computed from natural images using the classical \emph{grey world assumption}~\cite{GreyWorld}, or simply take a flat spectrum reference as we did here. 

\subsection{The trivial identity network}

The counter-example presented in this note is based on a trivial network architecture. 
Its output is the same as the input: for a color $C$, represented at the input by the array $R(C)$, the response $R'(C)$ is just: 
\begin{equation}
\label{eq:in}
    R'(R(C)) = I \cdot R(C) = R(C)
\end{equation}
This, clearly non-human, trivial architecture preserves whatever previous color representation coming from the sensors.
This trivial network is a good counter-example for the eventual human-like results because in the brain, the color representation in the retina certainly changes downstream~\cite{Shapley11,Shapley2019PhysiologyOC}.

\section{Experiments and Results}
\label{experiments}

As stated in the \emph{Methods} section, the conventional interpretation of the classical hue cancellation experiment can be questioned if one finds a counter example showing that human-like opponent valence curves may emerge for the classical choice of $\lambda_c$'s regardless of the post-retinal network architecture and color representation.
Moreover, refutation would be stronger if one finds that the human-like results are mainly obtained for the classical choice of $\lambda_c$'s while other choices lead to 
progressively different curves regardless of the input color representation space. 

According to this, we perform two sets of experiments:
(1) we look for counter examples with the classical hue cancellation lights using trivial identity networks working with different color representations (LMS, ATD and digital RGB).
(2) we consider a range of experiments with alternative cancellation lights different from the classical choice using the same trivial identity networks operating either in LMS, ATD or digital RGB.

\subsection{Counter examples in the classical setting}

\begin{figure}[t!]
\begin{center}
\includegraphics[width=12cm]{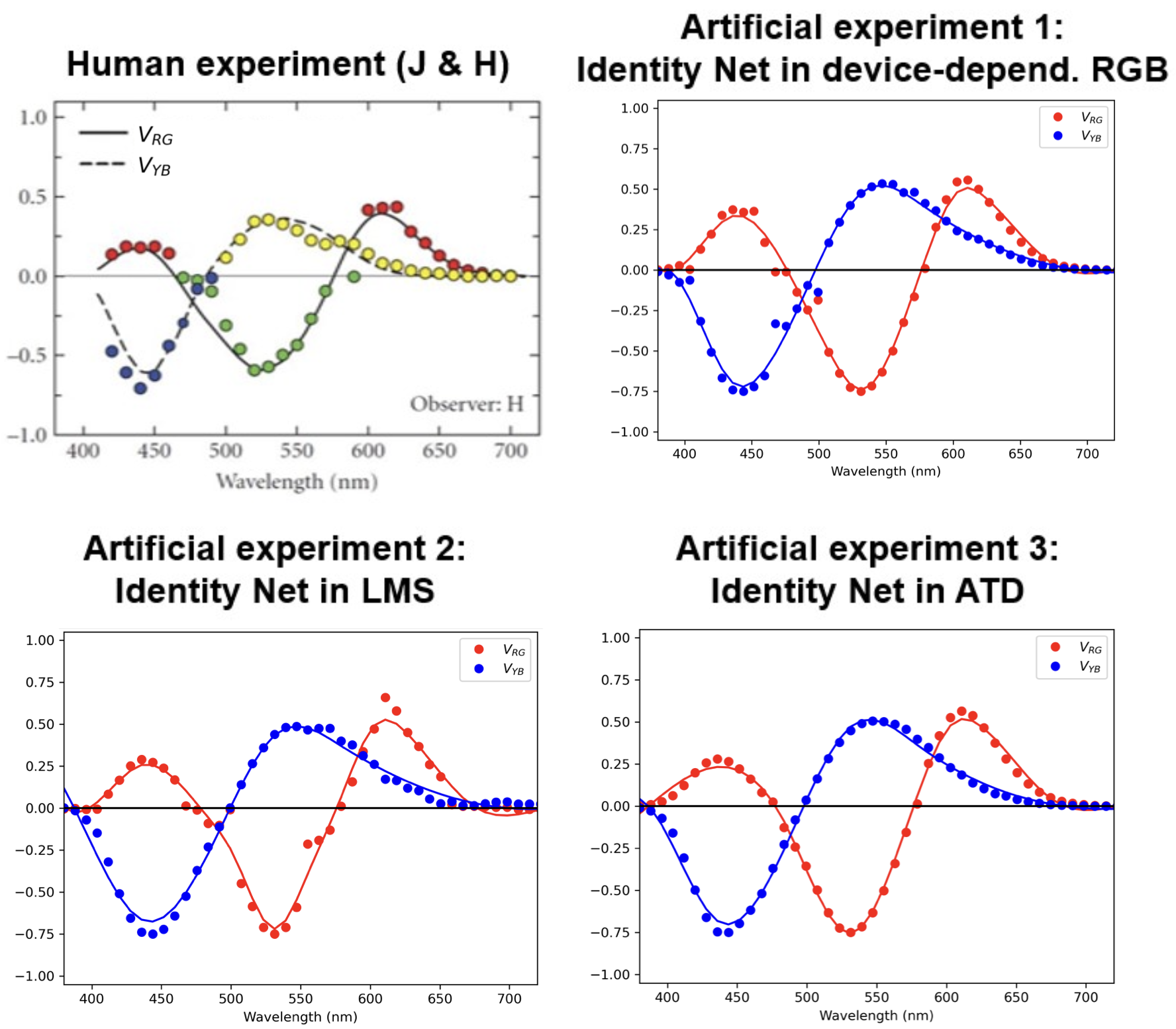}
\end{center}
\caption{Opponent curves for the trivial identity network operating in different color representation spaces. \label{results1}}
\end{figure}

In order to check the emergence of human-like curves in hue cancellation even with the trivial identity network, we perform three experiments assuming different \emph{input} representations $R$:
\begin{itemize}
    \item \textbf{Experiment~1:}~Identity network working in an arbitrary non-human color representation: a device-dependent digital RGB.
    \item \textbf{Experiment~2:}~Identity network working in a standard LMS cone space, as for instance~\cite{Stockman00}.
    \item \textbf{Experiment~3:}~Identity network working in a standard opponent space as for instance, the Jameson and Hurvich model~\cite{Jameson55_1,Capilla98}.
\end{itemize}

Note that the above three identity networks would correspond to color representations with quite different qualitative features: (a) if the input is digital RGB, the problem is solved by a system with wide-band overlapping all-positive spectral sensitivities (different from LMS) and compressive nonlinear response in the retina, (b) if the input are standard LMS tristimulus one has a purely linear LMS color code with all-positive sensitivities in the retina, and (c) if the input representation $R$ is an opponent system with an achromatic channel and two chromatic channels, the network is fed with a fundamentally different color coding.


Figure~\ref{results1} shows the results of these three hue cancellation experiments together with the experimental results for humans reported in~\cite{Jameson55_1}.

Appendix C shows that (1) the final matches make sense (found at the yellow-blue and red-green curves) and are close to perfect (almost zero difference after the addition of $w^\star_{\lambda_c}\!(\lambda) E_{\lambda_c}$), and (2) the difference minimization process with the different networks is remarkably similar.

The results show that all identity networks, regardless of the space where they operate, lead to similar hue cancellation curves, and these are remarkably similar to the human curves. 

\subsection{Alternative $\lambda_c$'s: control experiments and theoretical analysis}

The previous artificial experiments question the traditional interpretation of hue cancellation with the classical $\lambda_c$'s because not only opponent systems but also trichromatic systems lead to similar opponent results.
As anticipated above, the fortunate selection of the cancellation $\lambda_c$'s is \emph{somehow} biasing the matching towards the opponent curves.

In order to confirm that this is the case, we propose additional control experiments with artificial networks (experiments 4, 5 and 6), and we introduce a \emph{change-of-basis analogy} of the hue cancellation to understand the results. We show the predictions of this \emph{change-of-basis analogy} in the experiment~7:
\begin{itemize}
    \item \textbf{Experiment~4:}~Numerical results of hue cancellation for a range of $\lambda_c$'s away from the classical choice using the identity network working in a device-dependent digital RGB space.
    
    \item \textbf{Experiment~5:}~Numerical results of hue cancellation for a range of $\lambda_c$'s away from the classical choice  using the identity network working in a standard LMS space~\cite{Stockman00}.

    \item \textbf{Experiment~6:}~Numerical results of hue cancellation for a range of $\lambda_c$'s away from the classical choice using the identity network working in a standard ATD space~\cite{Jameson55_1,Capilla98}.
    
    \item \textbf{Experiment~7:}~Exhaustive exploration of (analytical) changes of basis that are similar to hue cancellation experiments for $\lambda_c$'s very different from the classical choice.
    
\end{itemize}

First, lets introduce the idea of the \emph{change-of-basis analogy} of the hue cancellation experiments, and then we present the results of experiments 4-6 together with the theory-based simulation (experiment 7). 

\begin{figure}[b!]
\begin{center}
\hspace{-0cm}\includegraphics[width=16cm]{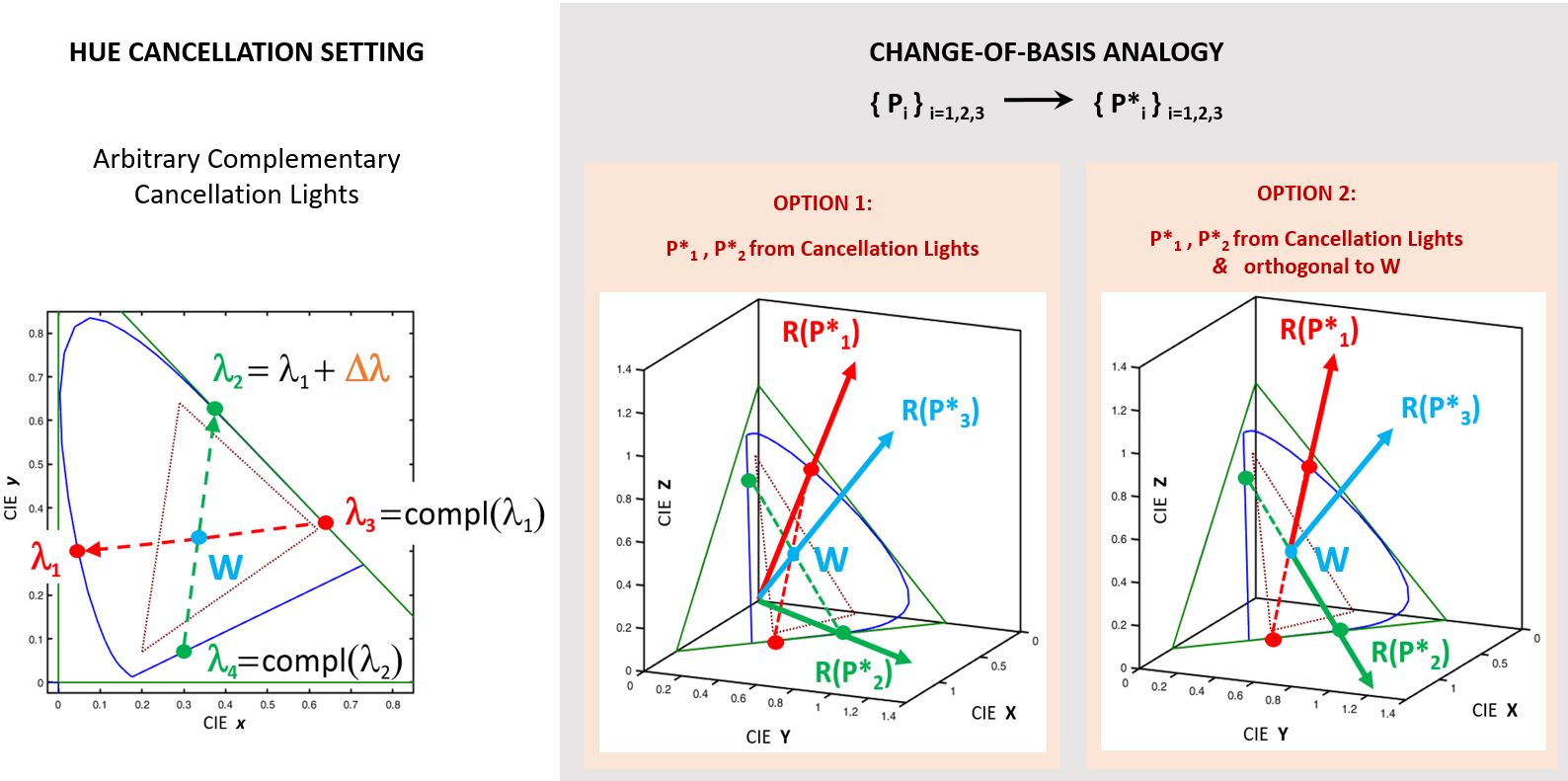}
\end{center}
\caption{The \emph{change-of-basis analogy:} Hue cancellation experiment as combination of vectors of a new basis. Note that the primaries $P^\star_i$ (based on the cancelling lights) are not related to the unknown primaries of the unknown representation $R'$. The primaries $P^\star_i$ (either in option 1 or 2) are just an \emph{artifice} to compute analytically the weights $w^\star_i(\lambda)$ from
the tristimulus values $R^\star_1(E_\lambda)$ and $R^\star_2(E_\lambda)$.
Given two arbitrary $\lambda_1$ and $\Delta \lambda$, the difference between \emph{option 1} and \emph{option 2} is that in the second the primaries $P^\star_1$ and $P^\star_2$ are taken to be orthogonal to the one that goes in the direction of the White, $P^\star_3 \propto W$, so that they convey \emph{less} information about brightness.}
\label{fig:surface_jesus1}
\end{figure}

Consider the case in which the cancellation lights are complementary in pairs. For instance, in Fig.~\ref{fig:surface_jesus1}, see the pair [$\lambda_1$, $\lambda_3$] and the pair formed by $\lambda_2$ and the magenta referred to as $\lambda_4$. 
In that situation, the determination of $w^\star_{\lambda_c}$ is equivalent to a change to a color basis where two of the primaries go in the directions of the pair of complementary wavelengths (e.g. the red and green vectors in Fig.~\ref{fig:surface_jesus1}). 
By choosing a third linearly-independent vector (e.g. in the direction of an achromatic color as the vector in blue perpendicular to the triangle of the chromatic diagram) one has a \emph{new basis} of the color space perfectly defined by the \emph{new} primaries, $P^\star_i$, with $i=1,2,3$.
These \emph{new} primaries are defined by their tristimulus vectors, $R(P^\star_i)$, in the basis of \emph{old} primaries, $P_i$, with $i=1,2,3$. They have chromatic coordinates $r(P^\star_i)$, and, as in every array of chromatic coordinates and tristimulus vectors, they are proportional: $R(P^\star_i) = \gamma_i  r(P^\star_i)$.

In this situation, taking $P_i$ as the input color representation (as in Fig.~1.b), hue cancellation with the four lights is analogous to a \emph{change-of-basis} from $P_i$ to $P^\star_i$. 
Therefore, looking for $w^\star_{\lambda_1}(\lambda)$ and $w^\star_{\lambda_2}(\lambda)$
is analogous to the computation of the tristimulus values of the monochromatic components of the equienergetic white $R^\star_1(E_{\lambda})$ and $R^\star_2(E_{\lambda})$.
Under this \emph{change-of-basis analogy}, the valence functions can be computed analytically from the color matching functions (the vectors $R(E_\lambda), \, \forall \, \lambda$), and the matrix $M_{PP^\star}$ that changes the vectors from the basis $P_i$ to the basis $P^\star_i$:
\begin{equation}
    R^\star(E_\lambda) = M_{PP^\star} \cdot R(E_\lambda)
    \label{change1}
\end{equation}
where, as in any standard change of basis~\cite{Stiles00}, the matrix is:
\begin{equation}\label{change2}
  M_{PP^\star} = \left(
    \begin{array}{ccc}
     R_1(P^\star_1) & R_1(P^\star_2) & R_1(P^\star_3) \\
     R_2(P^\star_1) & R_2(P^\star_2) & R_2(P^\star_3) \\
     R_3(P^\star_1) & R_3(P^\star_2) & R_3(P^\star_3) \\
    \end{array}
  \right)^{-1} = 
  \left(
    \begin{array}{ccc}
     \gamma_1^{-1}  & 0 & 0 \\
     0 & \gamma_2^{-1} & 0 \\
     0 & 0 & \gamma_3^{-1} \\
    \end{array}
  \right) \cdot
  \left(
    \begin{array}{ccc}
     r_1(P^\star_1) & r_1(P^\star_2) & r_1(P^\star_3) \\
     r_2(P^\star_1) & r_2(P^\star_2) & r_2(P^\star_3) \\
     r_3(P^\star_1) & r_3(P^\star_2) & r_3(P^\star_3) \\
    \end{array}
  \right)^{-1}
  \nonumber
\end{equation}

In this \emph{change-of-basis analogy} the hue cancellation valence functions are obtained from the color matching functions in the input representation transformed by the matrix in Eq.~\ref{change1}.  
Note that the weights $\gamma_i$ associated to the (arbitrary) length of the vectors, $R(P^\star_i)$, will scale each output $R^\star_i(E_\lambda)$. 
Therefore, despite the shape of the curves is fixed by the matrix of chromatic coordinates of the new basis, the global scale of the predicted functions can be varied via the length of the primaries. 
As a result, in the simulations using this analogy, given certain cancellation $\lambda_c$'s, the length of the basis vectors will be adjusted to obtain the best possible match between the predicted function and the classical curves of Jameson and Hurvich.

As explained in Appendix B, in the settings where the cancelling lights are not strictly complementary (as in the classical setting by Jameson and Hurvich) the curves can be obtained from alternative instrumental lights which are complementary. Then, the contribution of these instrumental lights always can be assigned back to the considered cancelling lights.
Therefore, (1) the classical setting can be understood using this \emph{change-of-basis analogy}, and (2) this analogy can be used to explore multiple combinations of axes $(\lambda_1, \lambda_3)$ and $(\lambda_2 = \lambda_1 + \Delta \lambda, \lambda_4)$. These configurations  can include the original experiment and also other, progressively different, alternatives. 

In the experiments 4-6 we execute artificial hue cancellation experiments with identity networks using complementary cancelling lights selected according to the \emph{change-of-basis analogy} described above. 
We explore a range of  $\lambda_1$ over the visible spectrum, and for each $\lambda_1$, we select $\lambda_2 = \lambda_1 + \Delta \lambda$ with a range of $\Delta \lambda$ so that $\lambda_2$ is still visible. Then, the 3rd and 4th cancellation lights are the complementary lights of $\lambda_1$ and $\lambda_2$.
Sometimes the complementary cancellation lights are purple-magenta, as in the arbitrary example of Fig.~\ref{fig:surface_jesus1}, but that is not a conceptual problem to apply the change-of-basis analogy.
We take the wavelengths in these control experiments along a uniform grid over the spectral space. 
The analytical solution of the change-of-basis analogy (Fig.~\ref{fig:surface_jesus1} and Eq.~\ref{change1}) can, of course, be used in this range of $\lambda_c$'s. Moreover, its analytical nature implies that one can efficiently sample the spectral space at higher rates. 
On top of the coarse regular grid shown below, we also perform the artificial hue cancellation at the configurations where the theory predicts better agreement with the opponent curves, which incidentally coincide with the wavelengths chosen in the classic experiment.

\begin{figure}[h!]
\begin{center}
\hspace{-2cm}
\includegraphics[width=18cm]{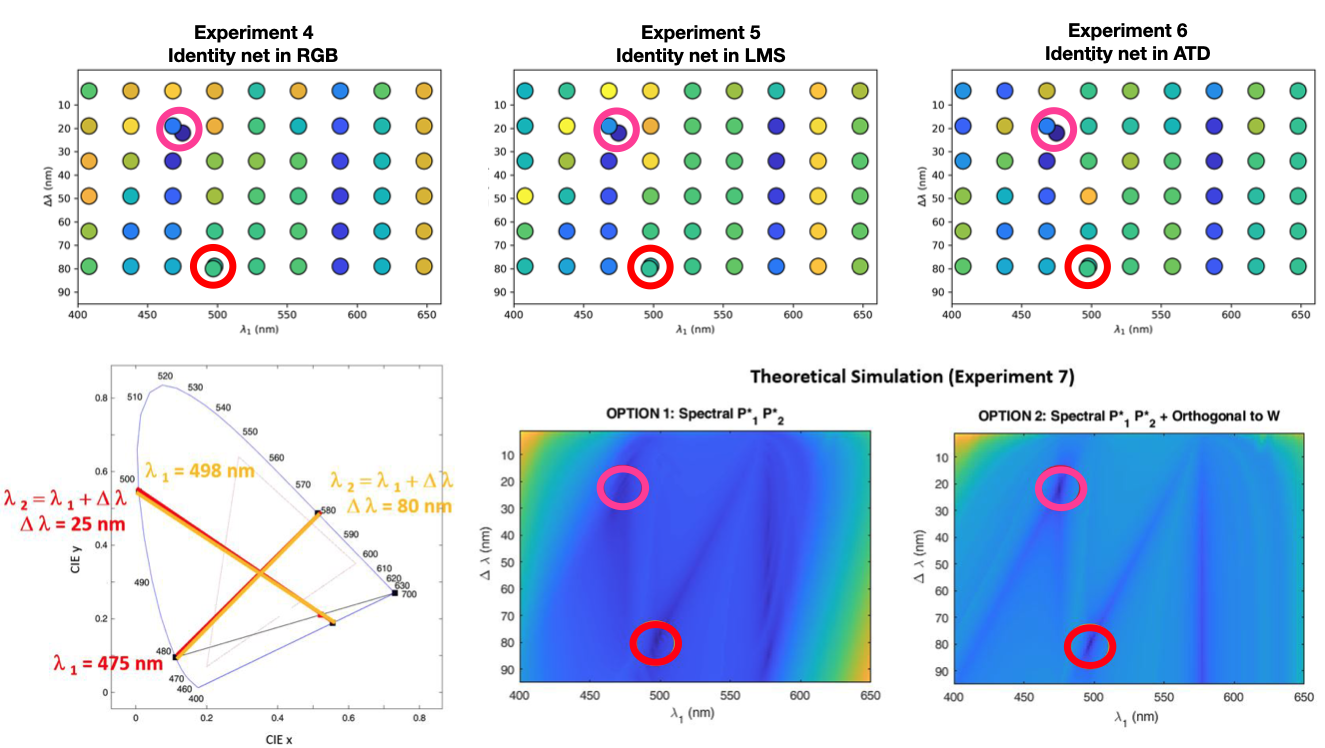}
\end{center}
\caption{Results of the control experiments (regular grid) together with the results in the original configuration (see the two dots off the regular grid).
\textbf{Top row} shows the errors of the experiments 4-6 with a blue-yellow colorbar scale where blue means low error (good reproduction of the human opponent curves) and yellow means high departure from the human result. The color code of the departure represents the Mean Squared Error between the human and the artificial curves. 
\textbf{Bottom row (right):} these surfaces represent the same kind of errors, with the same color code for the two options of the change-of-basis analogy.
The circles in red and magenta indicate the minima of the theoretical surfaces.
\textbf{Bottom row (left):} the chromatic diagram shows that the two minima found by the theoretical simulations actually correspond to the same choice of cancellation lights, and coincide with the classical setting (see appendix B for more information on the auxiliary magenta).}
\label{fig:surface_jesus2}
\end{figure}

For every considered configuration of cancellation lights we compute the cancellation (or valence) curves and we compute the departure from this result and the human curves of Jameson and Hurvich.  
Fig.~\ref{fig:surface_jesus2} shows the error of these predicted valence curves obtained either through the identity networks operating in different color spaces (experiments 4-6), or through the analytical change-of-basis analogy (experiment 7).

The results of experiments 4-7 stress the role of the choice of the cancellation lights in these experiments.
Note that \emph{all the error surfaces} have the same specific structure: 
\begin{itemize}

\item The theoretical surfaces of experiment 7 (which could be densely sampled since they are faster to compute) show two clear minima consistent with the setting selected in the classical experiment. The diagram shows that these two minima are actually equivalent.
Moreover, they display a clear pattern of secondary minima. The pattern is more distinct in the setting where the \emph{chromatic} primaries $P^\star_1$ and $P^\star_2$ are chosen to be orthogonal to the White.

\item The errors checked at the grid in the artificial hue cancellation experiments 4-6 are consistent with the theoretical surfaces despite the sampling grid is coarser. The reason for a coarser grid is merely computational\footnote{Each location  involves the estimation of the two valence curves at 50 $\lambda$'s. Therefore, it involves 50 hue cancellation experiments, i.e. 50 minimizations, one per $\lambda$ in the visible range}.
In some cases the deepest minimum is not in the classical point, but the difference is always very small, i.e. in the classical setting the artificial curves are also very similar to the human curves.

\item The artificial experiments lead to more marked differences between the agreement in the singular locations of small error (blueish points) and the rest. Note that the errors in the artificial experiments seem to increase faster as one goes away from the regions of small error. 

\end{itemize}

These results (which are consistent regardless of the use of trichomatic representations or opponent representations) suggest that the emergence of the classical curves is more linked to the selection of the cancellation lights than on the inner color representation $R'$. 

\section{Discussion}
\label{discussion}

\subsection{Summary of results}

When using trivial (identity) artificial networks in the classical hue cancellation setting, 
opponent red-green and yellow-blue valence functions emerge regardless of the actual color representation used by the networks (as long as it is a tristimulus representation or even tristimulus-like digital-RGB representations that include mild nonlinearities).

This suggests that these opponent curves do not inform us about the inner workings of the considered system, but about the properties of color mixtures in the tristimulus representations.
Given the fact that the mixture of opponent spectral cancellation lights is in the line between them in the chromatic diagram, changing the energy of these cancellation stimuli will always lead to displacements along these lines and hence, proper match with the grey reference (or proper hue cancellation) using the correct proportion of cancellation lights: humans and also trivial machines forced to use spectral (or quasi-spectral) cancellation lights would arrive to the same conclusion.

The reasoning is not as (analytically) obvious in nonlinear representations (as the digital-RGB) but results show that it follows the same trends, thus stressing the generality of the result.

The actual variation of the mixture when modifying the weights in the hue cancellation process only depends on the properties of the additive color mixture, and the path in the diagram is determined by the (classical) choice of the spectral cancellation lights, and not by the inner color representations.
Results suggest that a fortunate selection of the cancellation $\lambda_c$'s is \emph{somehow} biasing the matching towards the correct opponent curves.
If a range of alternative cancellation lights are considered, the results are progressively different from the classical opponent functions. 

With the classical $\lambda_c$'s, the different color representations only imply different metric spaces to compute the error in the match, but in absence of neural noise (or in moderate neural noise), this would mean minor variations in the result of the minimization, and hence one cannot rule out trichromatic LMS-like representations.

\subsection{Previous criticisms to hue cancellation experiments}

Certainly there are have been a number of well founded criticisms to the classical hue cancellation results.
For instance, \cite{Wandell95} makes this point: to what extent can we generalize from the valence measurements using monochromatic lights to other lights?. If the human behavior for polychromatic light does not follow from the behavior for monochromatic lights, then the data represents only an interesting (but non-generalizable) collection of observations.
In general, the linearity assumption is only an approximation \cite{Larimer74, Burns84, Ayama89, Chichilnisky95}. As a result, we need a more complete (nonlinear) model before we can apply the hue cancellation data to predict the opponent-colors appearance of polichromatic lights.
Other criticisms refer to overestimation of valence in certain spectral regions in hue cancellation versus other psychophysical methods~\cite{Ingling77,InglingTsou78,Ayama89}.

However, the problem implied by the systematic emergence of the opponent curves from the identity networks is different. It is not restricted to the linearity assumption. In fact, the systems with nets operating in the LMS or ATD spaces are linear by definition.
The emergence of the same result in two different (linear) trivial cases implies that the curves do not give a conclusive message about the inner working of the system.

\subsection{Emergence of human-like opponent curves in artificial systems}

Emergence of human-like behavior in artificial systems has been an inspiration for functional (or principled) explanations in theoretical neuroscience~\cite{Dayan01,Barlow59,Barlow01}.

In particular, due in part to the current success of artificial networks in vision tasks~\cite{AlexNet}, 
there is a growing interest to compare their behavior with humans~\cite{Bethge21,Geirhos1,Geirhos2} or with human-like models of traditional visual neuroscience~\cite{Martinez19,GomezVilla20,Li22,Hepburn22,Arash23}.

In this context, we set a low-level conventional psychophysics program to check the basic behavior of artificial networks in light of known basic human behavior~\cite{Bristol22,CIP_talk_22}.
In this context, to our surprise, our first experiments with artificial networks (with markedly non-human color representation) actually displayed human-like behavior in hue cancellation~\cite{CIP22}. 

That was the origin of this research because the emergence of human-like curves in hue cancellation in networks where opponency had not been built in 
(nor assumed in the training tasks) could have two implications:
\begin{itemize}

    \item \textbf{Hypothesis A:} On the positive side, it could imply that the considered tasks used to train the nets actually lead to human behavior in scenarios different from the training. 
    These evidences are interesting in the debate about the kind of tasks that may lead to human behavior. Note that certain tasks (e.g. assessing image quality or enhancing the retinal image), may lead to positive or negative results in reproducing human behavior depending on the architecture of the net. Consider examples in~\cite{Malo10,Martinez19} for the emergence of contrast nonlinearities, examples  in~\cite{Arash23,Li22} for the emergence of the Contrast Sensitivity Functions, or examples in~\cite{Hernandez23,Kumar22} for the visibility of distortions.
    
    \item \textbf{Hypothesis B:} On the negative side, it could also be that the experimental setting somehow forces the result. In this case the opponent curves would  not tell much about the inner color representation of the system, but about the selected \emph{opponent} spectral cancelling lights and about the properties of additive mixtures in tristimulus spaces. These elements (alien to the specific color coding in the network) could also explain the human-like opponent curves.

\end{itemize}

According to the results reported here, the second hypothesis seems the one that may be true.


\subsection{Implications in Visual Neuroscience}

Direct physiological recording of the opponent spectral sensitivity of cells~\cite{Devalois66,Derrington84} is (of course) the strongest indication of opponent color coding in the brain. 
However, following our results with trivial networks, the consistent emergence of the opponent curves in hue cancellation experiments suggests that other psychophysical techniques~\cite{KRAUSKOPF1982} may be more appropriate than hue cancellation to reveal the opponent mechanisms.
Similarly, our results suggest that indirect statistical arguments actually give stronger evidences in favour of opponent color coding than hue cancellation experiments. 
Statistical arguments are not limited to classical linear decorrelation~\cite{Buchsbaum1983TrichromacyOC,Ruderman:98}, but also include more recent, nonlinear measures of dependence~\cite{MacLeod03b,Laparra12,Gutmann14,Laparra15}.

\section*{Appendix A: Quasi-monochromatic spectrum  and cancellation lights}

\begin{figure}[h!]
\begin{center}
\includegraphics[width=\textwidth]{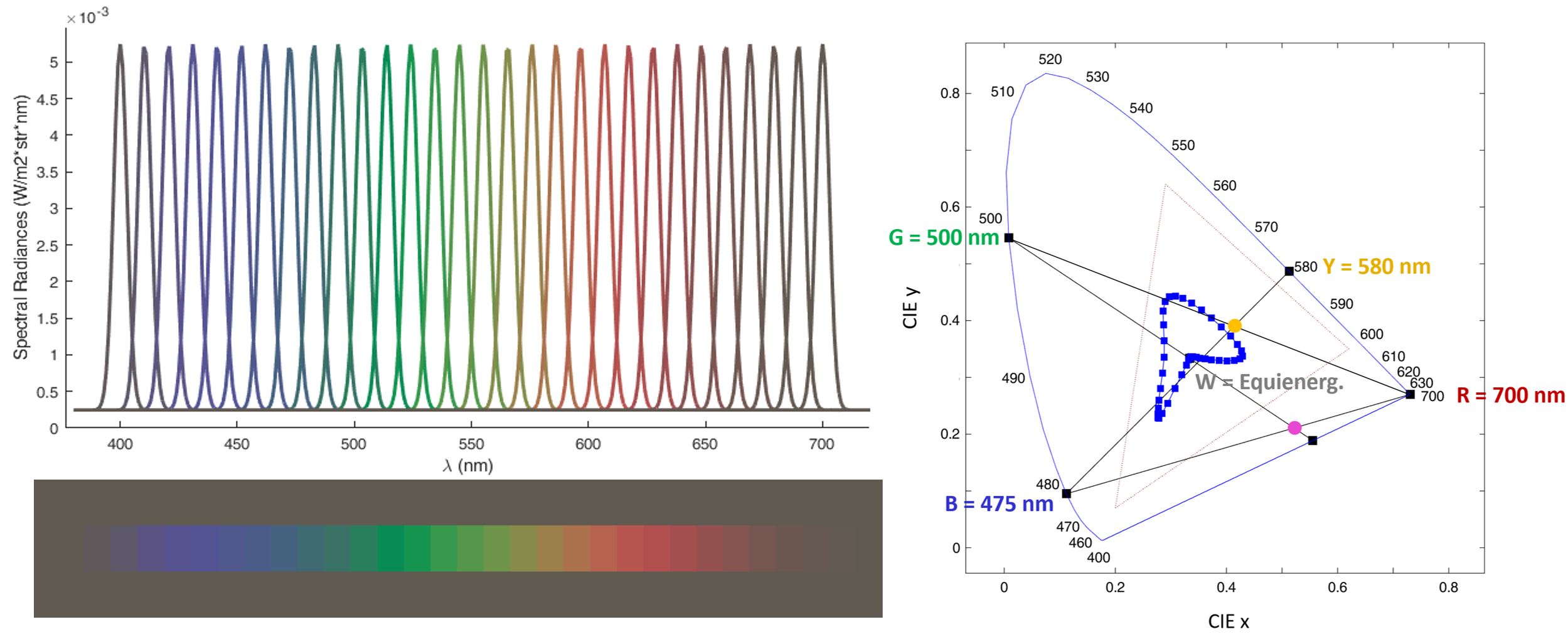}
\end{center}
\caption{Monochromatic cancellation lights and quasi-monochromatic approximation of the spectral locus. The auxiliary colors in \emph{yellow} and \emph{magenta} 
represent alternative methods to get the valence cancellation curves (see Appendix B) in case the complementary of some of the selected cancellation wavelengths is not a monochromatic stimulus (i.e. it is in the purple region) as is the case in the classical setting depicted here.
\label{appA}}
\end{figure}

Monochromatic lights live on the spectral locus of the color diagram. However, it is not possible to represent perfect monochromatic lights in digital values, so we used a quasi-monochromatic approximation to perform the experiments. To do that, we generate the quasi-monochromatic radiation as a narrow Gaussian spectral radiance of a determined height and width over a low-radiance equienergetic background. Fig.~\ref{appA} left shows the quasi-monochromatic spectrum generated for different lambdas. For the experiments we use a Gaussian height of $1.5\times10^{-3} W\cdot m^{-2}\cdot st^{-1}\cdot nm^{-1}$
over an equienergetic background of $0.5\times10^{-4} W\cdot m^{-2}\cdot st^{-1}\cdot nm^{-1}$
and a Gaussian width of $10 nm$.

Fig.~\ref{appA} right shows where the classical monochromatic lights live on the CIE 1931 color space (B, G, Y and R points) and the quasi-monochromatic reference wavelengths used in the experiments (inner blue points, which are inside the inner triangle that represent the color space that is possible to represent in digital values). The diagram also shows the equivalent opponent magenta marked by the red point in the B-R line that we used to combine the weights (see appendix B for more details).

\section*{Appendix B: valence functions from optimal $w^\star_{\lambda_c}\!(\lambda)$}

The result of our experimental settings with identity nets (as in the classical experiment) are four weights, $w^\star_{\lambda_c}\!(\lambda)$, obtained after solving Eq.~\ref{goal}. However, as in the classical experiment, we need to combine the four weights to obtain two curves, the \emph{red-green} and the \emph{yellow-blue} when the classical cancellation lights are used. The way to combine the weights depends on the cancellation lights. 

\subsection*{B.1 When the cancellation lights \emph{are} complementary}
Some cancellation stimuli can have \emph{complementary wavelengths}: for instance, in the conventional setting, $\lambda_{475}$ and $\lambda_{580}$ are approximately \emph{complementary} because their mixture can lead to a grey (approximately equal to the equienergetic white). The mixture of these two lights leading to the white can be obtained by solving the following equation:
\begin{equation}
    R(W) = \kappa_1 \cdot R(E_{475}) + \kappa_2 \cdot R(E_{580})
    \label{restriction1}
\end{equation}
Where $R(W)$ and $R(E_{\lambda})$ represent the tristimulus vectors of the white and the cancellation lights respectively, and $\kappa_{i}$ are the corresponding weights so that the sum of the two lights give monochromatic white.
Then, the corresponding cancellation weights (i.e. $w^\star_{475}$ and $w^\star_{580}$) are straightforward to mix because a positive increase in one of them can be compensated (in terms of hue) by a corresponding positive increase in the other with the corresponding $\kappa_i$ factors. With such same-sign increases, the mixture will remain at the same point in the chromatic diagram and hence the hue is not modified. As a result, these same-sign increments cancel. Similarly, weights of different sign in complementary $\lambda_c$'s contribute to the change of hue in the same way (moving the mixture in the same direction). Therefore, such opposite-sign increases should not cancel, but should be added in absolute value. In these opposite-sign cases, the resulting sign depends on the criterion taken to define the chromatic channel: for instance, if we decide to build a \emph{yellow-blue} channel (meaning positive values for long wavelengths and negative values for short wavelengths), the sum of modulus should be given a positive value when $w^\star_{475}<0$ and $w^\star_{580}>0$. In short, the yellow-blue valence function, $V_{\textrm{YB}}$, is:

\begin{equation}
    V_{\textrm{YB}} = \pm \kappa_1 \cdot \lvert w^\star_{475}\rvert \pm \kappa_2 \cdot \lvert w^\star_{580}\rvert
    \label{curve_no_magenta}
\end{equation}

where, the sign criterion we have just discussed above leads to these four cases:

\begin{equation}
\label{eq:signs}
\begin{cases}
    \,\,\, \textrm{if} \,\,\, w^\star_{475} \geq 0, w^\star_{580} \geq 0  \implies V_{\textrm{YB}} =  sign(w^\star_{475}-w^\star_{580}) \left| \kappa_1 \cdot w_{475}^* - \kappa_2 \cdot w_{580}^\star \right|, \\
    \,\,\, \textrm{if} \,\,\, w^\star_{475} < 0, w^\star_{580} < 0  \implies   V_{\textrm{YB}} = sign(w^\star_{580}-w^\star_{475}) \left| \kappa_1 \cdot w_{475}^* - \kappa_2 \cdot w_{580}^\star \right|, \\
    \,\,\, \textrm{if} \,\,\,  w^\star_{475} \geq 0, w^\star_{580} < 0 \implies V_{\textrm{YB}} = -\left( \kappa_1 \cdot \lvert w_{475}^* \rvert + \kappa_1 \cdot \lvert w_{580}^\star \rvert \right), \\
    \,\,\, \textrm{if} \,\,\, w^\star_{475} < 0, w^\star_{580} \geq 0 \implies V_{\textrm{YB}} = \kappa_1 \cdot \lvert w_{475}^* \rvert + \kappa_1 \cdot \lvert w_{580}^\star \rvert\\
\end{cases}
\end{equation}

The prescription is equivalent for any arbitrary pair of complementary cancellation $\lambda_c$. 
\newline


\subsection*{B.2 When the cancellation lights are not complementary}

In the case of the red-green channel, the complementary direction of the $\lambda_{500}$ is not in the direction of the $\lambda_{700}$. The actual complementary color is in the purple region. In that case, summation of $w^\star_{500}$ and $w^\star_{700}$ is not as straightforward because grey is not a sum of these cancellation lights.

\subsubsection*{Method 1: cancelling the reddish-greenish appearance (matching an auxiliary yellow instead of the white)}
The authors of the classical experiment, \cite{Jameson55_1} considered $\lambda_{700}$ and $\lambda_{500}$ as complementary because they weren't cancelling at white, but they were looking to cancel the reddish or greenish hue. This is equivalent to (\ref{restriction1}) but changing $W$ by an \emph{auxiliary yellow}, $\mathcal{Y}$, at the intersection of the YB line
with the line that connects the green $\lambda = 500 nm$ with the red $\lambda = 700 nm$.
See this line and the auxiliary yellow in the diagram of  Fig.~\ref{appA}, which can be obtained from this mixture:
\begin{equation}
    R(\mathcal{Y}) =
\kappa_3 \cdot R(E_{500}) + \kappa_4 \cdot R(E_{700})      \label{restriction2}
\end{equation}

Where we set the (arbitrary) luminance of this auxiliary Yellow as the sum of the luminance of $E_{500}$ and $E_{700}$, and $\kappa_{i}$ are the corresponding weights so that the sum of the two corresponding lights give a color which is neither red nor green. After that, we 
combine the obtained weights following the same sign criterion as in (\ref{eq:signs}).

\subsubsection*{Method 2: matching the white through an auxiliary magenta}
There is yet another way to solve the problem: in order to be able to cancel $\lambda_{500}$ to the white, we need to find its complementary, and we can also impose that it lies in the $BR$ line (magenta point in the diagram of Fig.~\ref{appA}) so that we can relate it with the other $\lambda_c$'s in use. We calculate this auxiliary \emph{magenta}, as $R(\mathcal{M}) = \alpha_{M1} \cdot R(E_{475}) + \alpha_{M2}\cdot R(E_{700})$, and we impose that it has the same energy as the other cancelling lights. We can consider, without loss of generality, that this magenta is complementary of $\lambda_{500}$ so that, when they are mixed with the appropriate weights, they generate the White. This magenta is only an artifice to get the red-green curve from the obtained $w^\star_i$; it has not been used in the optimization process. Its equivalent cancellation curve can be obtained via $w^\star_M = \alpha_{M1} \cdot w^\star_{475} + \alpha_{M2} \cdot w^\star_{700}$. Then, we can impose the White sum condition as before to get the corresponding weights $\kappa_i$:
\begin{equation}
    R(W) = \kappa_{M1} \cdot R(E_{500}) + \kappa_{M2} \cdot R(\mathcal{M})
    \label{restrictionM}
\end{equation}

Now we can obtain the red-green valence curve, $V_{\textrm{RG}}$, as a sum of $w^\star_{500}$ and $w^\star_{M}$ as follows (taking into account the same sign criteria stated in Eq.~\ref{eq:signs}):
\begin{align}
    V_{\textrm{RG}} &= \pm \kappa_{M1} \cdot \lvert w^\star_{500}\rvert \pm \kappa_{M2} \cdot \lvert w^\star_{M}\rvert = \pm \kappa_{M1} \cdot \lvert w^\star_{500}\rvert \pm \kappa_{M2} \cdot (\alpha_{M1}\cdot \lvert w^\star_{475}\rvert + \alpha_{M2}\cdot \lvert w^\star_{700}\rvert) = \nonumber \\ &= \pm \kappa_{M1} \cdot \lvert w^\star_{500}\rvert \pm \kappa_{M2}\cdot \alpha_{M1} \cdot \lvert w^\star_{475}\rvert \pm \kappa_{M2}\cdot \alpha_{M2}\cdot \lvert w^\star_{700}\rvert
    \label{curve_magenta}
\end{align}

By doing this calculation, we are using $w^\star_{475}$ to get the two curves, which is something that our algorithm has not taken into account. To avoid using the energy of $\lambda = 475$ nm twice, we must remove from $V_{\textrm{RG}}$ the amount of $w^\star_{475}$ that we used in $V_{\textrm{YB}}$. Doing so,  Eq.~\ref{curve_magenta} becomes:
\begin{equation}
    V_{\textrm{RG}} = \pm \kappa_{M1} \cdot \lvert w^\star_{500}\rvert \pm (\kappa_{M2}\cdot \alpha_{M1} - \kappa_1)\cdot \lvert w^\star_{475}\rvert \pm \kappa_{M2}\cdot \alpha_{M2}\cdot \lvert w^\star_{700}\rvert
    \label{curve_magenta_good}
\end{equation}
 
\begin{figure}[t!]
\begin{center}
\includegraphics[width=12cm]{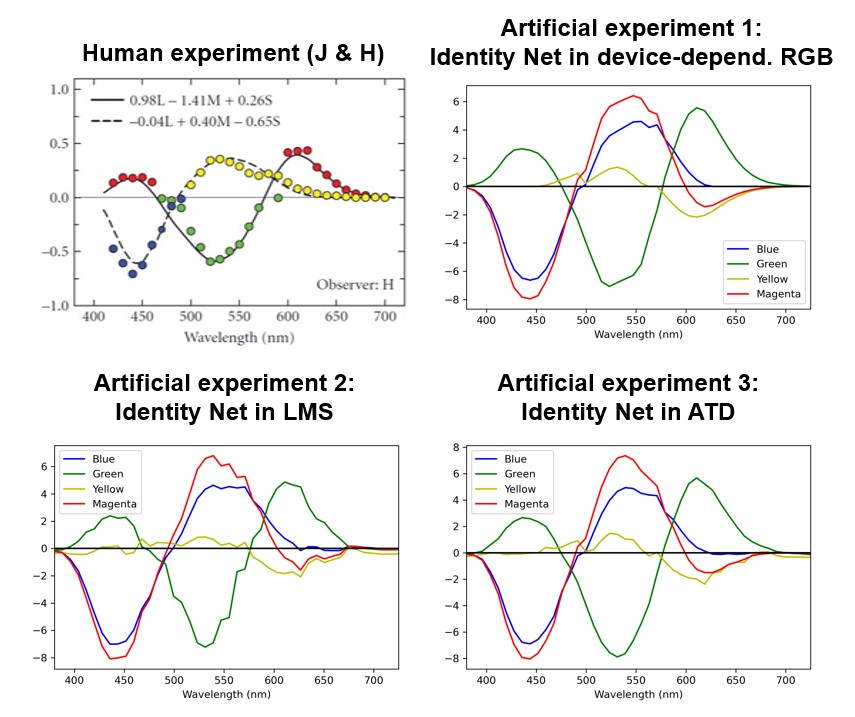}
\end{center}
\caption{Optimal weights, $w_i^\star(\lambda)$, of the classical cancellation lights for the trivial identity network operating in different photoreceptor color spaces. 
Here we show the magenta curve (built from the blue curve and the red curve) that can be directly subtracted from the green curve to obtain $V_{\textrm{RG}}$\label{results2}}
\end{figure}

Fig.~\ref{results2} right shows the $\lambda_{475}$ (blue), $\lambda_{580}$ (yellow), $\lambda_{500}$ (green) and auxiliary-magenta curves that are summed to give the yellow-blue and red-green curves.

The procedure described here can be applied to other choices of cancelling $\lambda_c$'s.
When exploring the whole range of possible cancelling $\lambda_c$ to simulate the hue cancellation experiment in situations beyond the conventional choice of cancelling lights, we always compute first the \emph{complementary} curves (one or two) when possible and then, when necessary, compute the complementary of $\lambda_c$ with the red or blue extremes to get the last curve. Note that we always use $\lambda_{700}$ when only one component lies in the purple line, but we use both $\lambda_{400}$ and $\lambda_{700}$ when there are two.

Finally, a note on the scaling of the valence curves. The shape of the curves and their relative scale determine how the matchings are made for each $\lambda$.
According to the change-of-basis analogy in Eq.~\ref{change1}, the scale of the curves is associated to the \emph{arbitrary} length of the associated primaries $P^\star_i$. Therefore, once the minimization is finished, we keep the spectral shape constant and we look for the optimal lengths of $P^\star_i$ to obtain the best match to the human-opponent curves.

\section*{Appendix C: Visualization of hue cancellation matches with classical $\lambda_c$'s}

\begin{figure}[h!]
\begin{center}
\includegraphics[width=\textwidth]{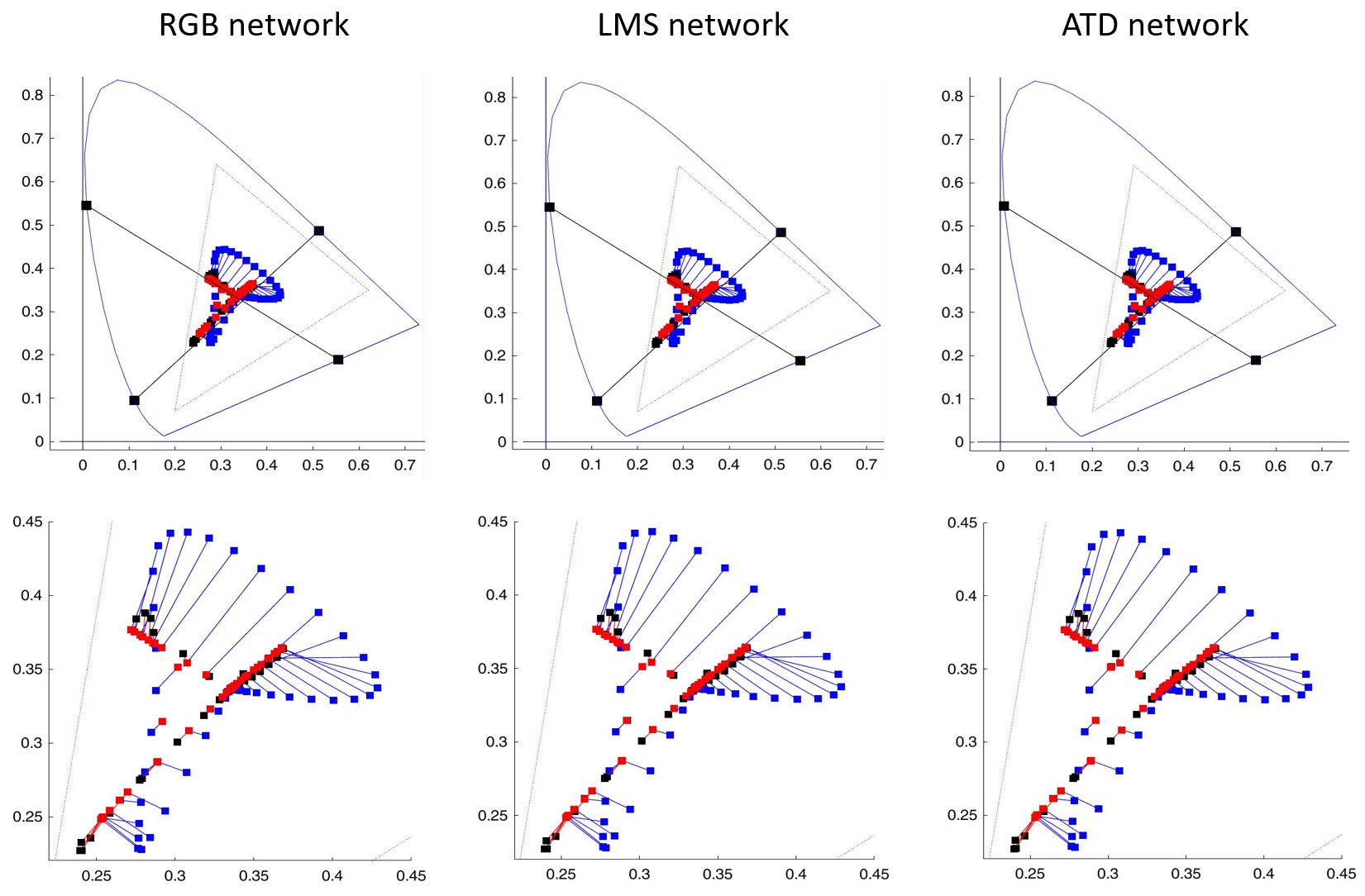}
\end{center}
\caption{Visualization of the hue cancellation solutions achieved by the networks considered in the experiments with the classical $\lambda_c$'s. The stimuli in blue are the original quasi-monochromatic stimuli (the initial state before the addition of $E_{\lambda_c}$). The red and the black stimuli correspond to the colors of the spectral and the reference modified with the addition of the optimal $E_{\lambda_c}$. 
Blue and black lines represent the initial and final differences before and after the optimization process.\label{appD}}
\end{figure}

It is important to check if the algorithm we used to minimize the distance has converged. In Fig.~\ref{appD} we represent the hue cancellation solutions after solving Eq.~\ref{goal} for experiments 1-3. Blue points represent the initial quasi-monochromatic stimuli, before the addition of the cancelling lights (i.e, $w^\star_{i} = 0$). Black and red points represent the colors of the spectral and reference modified with the addition of the optimal $E_{\lambda_c}$ founded by Eq.~\ref{goal}. We find that independently of the color representation, identity network gets the match at the directions determined by the selected $\lambda_c$'s in a very consistent way. Interestingly, the \emph{red-green} axis consistent with the magenta complementary of $\lambda_c=500$ nm was not imposed in any way because the minimization was done by modifying the energy of $\lambda_c=700$ nm.
Of course (as in any learning process prone to errors due to early stopping), the networks do not find the absolute minimum (in a perfect match the difference between the red and the black stimuli should be zero). However, the final differences (black lines) are substantially smaller than the initial differences (blue lines).

\section*{Acknowledgments}

The authors thank interesting discussions on the preliminary results~\cite{CIP22} that lead to this research with A. Parraga, A. Akbarinia, J. Vazquez-Corral, X. Otazu, M. Bertalmío, F. Wichmann, and particularly, V. Laparra. 
This work was supported in part by MICIIN/FEDER/UE under Grant PID2020-118071GB-I00 and PDC2021-121522-C21, in part by Spanish MIU under Grant FPU21/02256 and in part by Generalitat Valenciana under Projects GV/2021/074, CIPROM/2021/056 and CIAPOT/2021/9. Some computer resources were provided by Artemisa, funded by the European Union ERDF and Comunitat Valenciana as well as the technical support provided by the Instituto de Física Corpuscular, IFIC (CSIC-UV).


\medskip
\printbibliography

@article{Jameson57,
  title={An opponent-process theory of color vision.},
  author={Leo M. Hurvich and Dorothea Jameson},
  journal={Psychological review},
  year={1957},
  volume={64, Part 1 6},
  pages={384-404}
}

@misc{Hernandez23,
  doi = {10.48550/ARXIV.2302.13345},
  author = {Hernández-Cámara, Pablo and Vila-Tomás, Jorge and Laparra, Valero and Malo, Jesús},
  title = {Analysis of Deep Image Quality Models},
  publisher = {arXiv},
  year = {2023},
}

@article{Kumar22,
title={Do better ImageNet classifiers assess perceptual similarity better?},
author={Manoj Kumar and Neil Houlsby and Nal Kalchbrenner and Ekin Dogus Cubuk},
journal={Trans. Mach. Learn. Res.},
year={2022},
url={https://openreview.net/forum?id=qrGKGZZvH0},
}

@article{Barlow59,
  author={Barlow, H.B.},
  title={Sensory mechanisms, the reduction of redundancy, and intelligence},
  journal={Proc. of the Nat. Phys. Lab. Symposium on the Mechanization of Thought Process},
  number={10},
  pages={535-539},
  year={1959},
}

@book{Dayan01,
 author = {Dayan, Peter and Abbott, L. F.},
 title = {Theoretical Neuroscience: Computational and Mathematical Modeling of Neural Systems},
 year = {2005},
 isbn = {0262541858},
 publisher = {The MIT Press},
 ADDRESS = "Boston, MA"}

@ARTICLE{Barlow01,
	author = {Barlow, H.},
	title = {Redundancy reduction revisited},
	journal = {Network: Comp. Neur. Syst.},
	year = {2001},
        volume = {12},
	number = {3},
	pages = {241-253}
}

@inproceedings{AlexNet,
 author = {Krizhevsky, Alex and Sutskever, Ilya and Hinton, Geoffrey E},
 booktitle = {Advances in Neural Information Processing Systems},
 editor = {F. Pereira and C.J. Burges and L. Bottou and K.Q. Weinberger},
 pages = {},
 publisher = {Curran Associates, Inc.},
 title = {ImageNet Classification with Deep Convolutional Neural Networks},
 url = {https://proceedings.neurips.cc/paper/2012/file/c399862d3b9d6b76c8436e924a68c45b-Paper.pdf},
 volume = {25},
 year = {2012}
}

@inproceedings{Geirhos1,
title={ImageNet-trained {CNN}s are biased towards texture; increasing shape bias improves accuracy and robustness.},
author={Robert Geirhos and Patricia Rubisch and Claudio Michaelis and Matthias Bethge and Felix A. Wichmann and Wieland Brendel},
booktitle={International Conference on Learning Representations},
year={2019},
url={https://openreview.net/forum?id=Bygh9j09KX},
}

@ARTICLE{Geirhos2,
	author = {Geirhos, R. and Jacobsen, JH. and Michaelis, C. and Zemel, R. and Brendel, W. Bethge, M. and Wichmann, FA.},
	title = {Shortcut learning in deep neural networks},
	journal = {Nature Machine Intelligence},
	year = {2020},
    volume = {2},
	pages = {665–673}
}

@article{Bethge21,
    author = {Funke, Christina M. and Borowski, Judy and Stosio, Karolina and Brendel, Wieland and Wallis, Thomas S. A. and Bethge, Matthias},
    title = "{Five points to check when comparing visual perception in humans and machines}",
    journal = {Journal of Vision},
    volume = {21},
    number = {3},
    pages = {16-16},
    year = {2021},
    month = {03},
    issn = {1534-7362},
    doi = {10.1167/jov.21.3.16},
    url = {https://doi.org/10.1167/jov.21.3.16},
    eprint = {https://arvojournals.org/arvo/content\_public/journal/jov/938520/i1534-7362-21-3-16\_1615892377.49847.pdf},
}

@article{GomezVilla20,
  title = {Color illusions also deceive {CNNs} for low-level vision tasks: Analysis and implications},
  author = {Gomez-Villa, A. and Martin, A. and Vazquez, J. and Bertalm\'io, M. and Malo, J.},
  journal = {Vision Research},
  volume = {176},
  issue = {11},
  pages = {156-174},
  year = {2020},
}

@article{Li22,
    author = {Li, Qiang and Gomez-Villa, Alex and Bertalmío, Marcelo and Malo, Jesús},
    title = "{Contrast sensitivity functions in autoencoders}",
    journal = {Journal of Vision},
    volume = {22},
    number = {6},
    year = {2022},
    doi = {10.1167/jov.22.6.8},
}

@inproceedings{Hepburn22,
title={On the relation between statistical learning and perceptual distances},
author={Alexander Hepburn and Valero Laparra and Raul Santos-Rodriguez and Johannes Ball{\'e} and Jesus Malo},
booktitle={International Conference on Learning Representations},
year={2022},
url={https://openreview.net/forum?id=zXM0b4hi5_B}
}

@article {Arash23,
	author = {Akbarinia, Arash and Morgenstern, Yaniv and Gegenfurtner, Karl R.},
	title = {Contrast Sensitivity Function in Deep Networks},
	year = {2023},
	doi = {10.1101/2023.01.06.523034},
	journal = {bioRxiv}
}

@InProceedings{Bristol22,
author="Vila-Tomás, J.
and Hernández-Camara, P.
and Malo, J.",
editor="Santos, R.",
title="A Psychophysical Turing Test for Artificial Networks devoted to Vision",
booktitle="Workshop on Evaluating Artificial Intelligence",
year="2022",
publisher="Bristol Univ. Dept. Eng. Math,",
address="Bristol, UK",
}

@InProceedings{CIP_talk_22,
author="Hernández-Camara, P. and 
Vila-Tomas, J. and 
and Malo, J.",
editor="Parraga, A. and Otazu, X.",
title="A visual psychophysics decalogue to assess the human nature of artificial networks",
booktitle="Workshop on Deep Learning in Vision Science",
year="2022",
publisher="Univ. Barcelona, Iberian Conf. Percept.",
address="Barcelona, Spain",
}

@InProceedings{CIP22,
author="Vila-Tomás, J. and  Hernández-Camara, P. and Li, Q. and A. Hepburn, V. Laparra & J. Malo ",
editor="Parraga, A. and Otazu, X.",
title="Basic psychophysics of deep networks trained to reproduce segmentation, maximum differentiation and subjective distortions",
booktitle="Workshop on Deep Learning in Vision Science",
year="2022",
publisher="Univ. Barcelona, Iberian Conf. Percept.",
address="Barcelona, Spain",
}

@article{DeValois66,
title={Analysis of response patterns of {LGN} cells},
  author={DeValois, RL. and Abramov, I. and Jacobs, GH.},
  journal={J. Opt. Soc. Am.},
  year={1966},
  volume={56},
  pages={966-977}
}

@article{Ingling77,
title={The spectral sensitivity of the opponent-color channels},
  author={Ingling, Carl},
  journal={Vis. Res.},
  year={1977},
  volume={17},
  pages={1083-1089}
}

@article{Shapley11,
  title={Color in the Cortex—single- and double-opponent cells},
  author={Robert Shapley and Michael Hawken},
  journal={Vision Res.},
  year={2011},
  volume={51(7)},
  pages={701–717}
}

@BOOK{Stiles00,
  AUTHOR = "Wyszecki, G and Stiles, WS.",
  TITLE  = "Color Science: Concepts and Methods, Quantitative Data and Formulae",
  PUBLISHER = "John Wiley \& Sons",
  YEAR = 2000,
  ADDRESS = "New Jersey"}

@article{Jameson55_1,
author = {Dorothea Jameson and Leo M. Hurvich},
journal = {J. Opt. Soc. Am.},
number = {7},
pages = {546--552},
publisher = {Optica Publishing Group},
title = {Some Quantitative Aspects of an Opponent-Colors Theory. I. Chromatic Responses and Spectral Saturation},
volume = {45},
month = {Jul},
year = {1955},
url = {http://opg.optica.org/abstract.cfm?URI=josa-45-7-546},
doi = {10.1364/JOSA.45.000546},
}

@InProceedings{GreyWorld,
author="Finlayson, G.D.
and Schiele, B.
and Crowley, J.L.",
editor="Burkhardt, H.
and Neumann, B.",
title="Comprehensive colour image normalization",
booktitle="Computer Vision --- ECCV'98",
year="1998",
publisher="Springer",
address="Berlin, Heidelberg",
pages="475--490",
}

@article{Capilla98,
	doi = {10.1088/0150-536x/29/5/003},
	url = {https://doi.org/10.1088/0150-536x/29/5/003},
	year = 1998,
	month = {oct},
	publisher = {{IOP} Publishing},
	volume = {29},
	number = {5},
	pages = {324--338},
	author = {P Capilla and J Malo and M J Luque and J M Artigas},
	title = {Colour representation spaces at different physiological levels: a comparative analysis},
	journal = {Journal of Optics}
}

@article{Malo10,
  title={Psychophysically tuned divisive normalization approximately factorizes the PDF of natural images},
  author={Malo, J. and Laparra, V.},
  journal={Neural computation},
  volume={22},
  number={12},
  pages={3179--3206},
  year={2010},
}

@misc{Colorlab02,
  author = {Malo, J. and  Luque,  M.J.},
  title = {{ColorLab: A Matlab Toolbox for Color Science and Calibrated Color Image Processing}},
  howpublished = "\url{http://isp.uv.es/code/visioncolor/colorlab.html}",
  year = {2002},
}

@INBOOK{Stockman11,
  author =       {Stockman, A. and Brainard, D.H.},
  editor =       {Bass, M.},
  title =        {OSA Handbook of Optics (3rd. Ed.)},
  chapter =      {Color vision mechanisms},
  pages =        {147--152},
  publisher =    {McGraw-Hill},
  year =         {2010},
  address =      {NY},
}

@INBOOK{Shevell04,
  author =       {Knoblauch, K. and Shevell, S.K.},
  editor =       {Chalupa, L. M. and Werner, J.S.},
  title =        {The visual neurosciences},
  chapter =      {57. Color Appearance},
  pages =        {892-907},
  publisher =    {MIT Press},
  year =         {2004},
  volume =       {1},
}

@book{Fairchild13,
  title={Color Appearance Models},
  author={Fairchild, M.D.},
  series={The Wiley-IS\&T Series in Imaging Science and Technology},
  year={2013},
  publisher={Wiley}
}

@BOOK{Wandell95,
  AUTHOR = "B.A. Wandell",
  TITLE  = "Foundations of Vision",
  PUBLISHER = "Sinauer Assoc. Publish.",
  YEAR = 1995,
  ADDRESS = "Massachusetts"}

@article{Psytoolbox97,
author = {Brainard, D. H.},
journal = {Spatial Vision},
number = {10},
pages = {433-436},
title = {The Psychophysics Toolbox},
year = {1997},
url = {https://www.psychopy.org/},
}

@article{Stockman00,
title = "The spectral sensitivities of the middle- and long-wavelength-sensitive cones derived from measurements in observers of known genotype",
journal = "Vision Research",
volume = "40",
number = "13",
pages = "1711 - 1737",
year = "2000",
author = "Stockman, A. and Sharpe, L.T."
}

@article{InglingTsou78,
author = {Carl R. Ingling  and Phillip W. Russell  and Mark S. Rea  and Brian H.-P. Tsou },
title = {Red - Green Opponent Spectral Sensitivity: Disparity Between Cancellation and Direct Matching Methods},
journal = {Science},
volume = {201},
number = {4362},
pages = {1221-1223},
year = {1978},
doi = {10.1126/science.201.4362.1221},
URL = {https://www.science.org/doi/abs/10.1126/science.201.4362.1221},
eprint = {https://www.science.org/doi/pdf/10.1126/science.201.4362.1221}}

@article{KRAUSKOPF1982,
title = {Cardinal directions of color space},
journal = {Vision Research},
volume = {22},
number = {9},
pages = {1123-1131},
year = {1982},
issn = {0042-6989},
doi = {https://doi.org/10.1016/0042-6989(82)90077-3},
url = {https://www.sciencedirect.com/science/article/pii/0042698982900773},
author = {John Krauskopf and David R. Williams and David W. Heeley}
}

@article{Derrington84,
author = {Derrington, A M and Krauskopf, J and Lennie, P},
title = {Chromatic mechanisms in lateral geniculate nucleus of macaque.},
journal = {The Journal of Physiology},
volume = {357},
number = {1},
pages = {241-265},
year = {1984}
}

@article{Ruderman:98,
author = {Daniel L. Ruderman and Thomas W. Cronin and Chuan-Chin Chiao},
journal = {J. Opt. Soc. Am. A},
number = {8},
pages = {2036--2045},
publisher = {Optica Publishing Group},
title = {Statistics of cone responses to natural images: implications for visual coding},
volume = {15},
month = {Aug},
year = {1998},
url = {http://opg.optica.org/josaa/abstract.cfm?URI=josaa-15-8-2036},
doi = {10.1364/JOSAA.15.002036},
}

@article{Gutmann14,
  title={Spatio-chromatic adaptation via higher-order canonical correlation analysis of natural images},
  author={Gutmann, M. U. and Laparra, V. and Hyv{\"a}rinen, A. and Malo, J.},
  journal={{PloS} {ONE}},
  volume={9},
  number={2},
  pages={e86481},
  year={2014},
  publisher={Public Library of Science}
}

@incollection{MacLeod03b,
    author = {MacLeod, D. and von der Twer, T.},
    title = {The pleistochrome: optimal opponent codes for natural colors},
    booktitle = {Color Perception: From Light to Object},
    address = {Oxford, UK},
    editor = {Heyer, D. and Mausfeld, R.},
    publisher = {Oxford Univ. Press},
    year = {2003}
}

@article{Buchsbaum1983TrichromacyOC,
  title={Trichromacy, opponent colours coding and optimum colour information transmission in the retina},
  author={Gershon Buchsbaum and Allan Gottschalk},
  journal={Proceedings of the Royal Society of London. Series B. Biological Sciences},
  year={1983},
  volume={220},
  pages={113 - 89}
}

@article{Shapley2019PhysiologyOC,
  title={Physiology of Color Vision in Primates},
  author={Robert Shapley},
  journal={Oxford Research Encyclopedia of Neuroscience},
  year={2019}
}

@article{Martinez19,
  author={Martinez, M. and Bertalm\'io, M and Malo, J.},
  title={In Praise of Artifice Reloaded: Caution with Natural Image Databases in Modeling Vision},
  journal={Front. Neurosci. doi: 10.3389/fnins.2019.00008},
  doi = {doi: 10.3389/fnins.2019.00008},
  year={2019},
}

@article{Laparra15,
  title={Visual aftereffects and sensory nonlinearities from a single statistical framework},
  author={Laparra, V. and Malo, J.},
JOURNAL={Frontiers in Human Neuroscience},
VOLUME={9},
PAGES={557},
YEAR={2015},
URL={http://journal.frontiersin.org/article/10.3389/fnhum.2015.00557},
DOI={10.3389/fnhum.2015.00557},
}

@article{Larimer74,
author = {Larimer, J.},
journal = {Vis. Res.},
volume = {14},
number = {11},
pages = {1127-40},
title = {Opponent-process additivity~I: red-green equilibria},
year = {1974},
doi = {10.1016/0042-6989(74)90209},
}

@article{Burns84,
	author = {S.A. Burns and A.E. Elsner and J. Pokorny and V.C. Smith},
	journal = {Vision Research},
	number = {5},
	pages = {479-489},
	title = {The abney effect: Chromaticity coordinates of unique and other constant hues},
	volume = {24},
	year = {1984}}

@article{Ayama89,
	author = {Miyoshi Ayama and Mitsuo Ikeda},
	journal = {Vision Research},
	number = {9},
	pages = {1233-1244},
	title = {Dependence of the chromatic valence function on chromatic standards},
	volume = {29},
	year = {1989}}

@article{Chichilnisky95,
	author = {EJ {Chichilnisky}},
	journal = {Standford Univ. PhD.},
	title = {Perceptual measurements of neural computation in color appearance},
	year = {1995}}

@ARTICLE{Laparra12,
  author={V. {Laparra} and S. {Jiménez} and G. {Camps} and J. {Malo}},
  journal={Neural Computation},
  title={Nonlinearities and Adaptation of Color Vision from Sequential Principal Curves Analysis},
  year={2012},
  volume={24},
  number={10},
  pages={2751-2788},
  doi={10.1162/NECO_a_00342}}

\end{document}